\documentclass[a4paper,11pt]{article}
\usepackage{pos}

\usepackage{amsmath,amssymb}
\usepackage{subcaption}
\usepackage{float}
\usepackage{doi}   

\title{Spatial String Tension at High Temperatures\\and Quantitative Tests
of Dimensionally Reduced Effective Theories}

\author*[a]{Vishal Rao}
\author[b]{Peter Petreczky}
\author[a]{Prasad Hegde}

\affiliation[a]{Center for High Energy Physics, Indian Institute of Science,\\
               Bangalore 560012, India}
\affiliation[b]{Physics Department, Brookhaven National Laboratory,\\
               Upton, NY 11973, USA}

\emailAdd{Vishalr@iisc.ac.in}
\emailAdd{petreczk@bnl.gov}
\emailAdd{prasadhegde@iisc.ac.in}

\abstract{%
We calculate the spatial string tension in 2+1 flavour QCD in $(3{+}1)$
dimensions within a temperature range of $[166\,\mathrm{MeV},\,1000\,\mathrm{MeV}]$
using spatial Wilson loops with HYP smearing.
We use the Highly Improved Staggered Quark action for fermions and the
tree-level Symanzik improved gauge action for gluons at two lattice
spacings corresponding to temporal extents $N_\tau = 8$ and $10$.
We then compare our results with dimensionally reduced effective theories
at high temperatures (EQCD and MQCD) to test the onset of dimensional
reduction in QCD.}

\FullConference{%
The $42^{\mathrm{nd}}$ International Symposium on Lattice Field Theory,
Lattice 2025,\\
$2^{\mathrm{nd}}$--$8^{\mathrm{th}}$ November 2025 @ TIFR Mumbai, India.}

\begin{document}
\maketitle

\section{Introduction}
Above the pseudo-critical temperature $T_{pc} \simeq 156\,\mathrm{MeV}$~\cite{HotQCD2019},
QCD undergoes a crossover to the quark-gluon plasma (QGP) phase.
An accurate picture of the QGP is far from the free gas of quarks and gluons even at temperatures much higher than $T_{pc}$.  
At finite temperature, the Euclidean time direction is compactified to a circle
of circumference $\beta = 1/T$, giving the partition function
$Z = \mathrm{Tr}(e^{-\beta H})$.
The periodic boundary conditions for bosons, anti-periodic for fermions generate Matsubara frequencies $2\pi n T$ and $(2n{+}1)\pi T$ respectively.
These modes can be sequentially integrated out to get effective field theories like EQCD and MQCD. 
The Linde problem \cite{Linde1980} tells us that the magnetic sector is non-perturbative for $r > 1/(g^2 T)$, which is also evident from expansion parameter of MQCD ($g^2T/p \sim 1$) for modes of momenta $p \sim g^2T$. 
A direct physical consequence is that even inside the QGP, magnetic gluons
remain confined.
Spatial Wilson loops (loops lying entirely in the three spatial directions) continue to obey an area law above $T_{pc}$, with the coefficient
defining the spatial string tension $\sigma_s$.
Unlike the ordinary zero-temperature string tension, which vanishes above $T_{pc}$,
$\sigma_s$ actually grows with temperature, tracking the magnetic mass scale $g^2T$.
The question is: at what temperature does the full $(3{+}1)$-dimensional QCD
reduce to the simpler effective theories that govern its magnetic sector?
The spatial string tension is an ideal observable to answer this, because
it exists in all three theories (full 4D QCD, EQCD, and MQCD).
Earlier work~\cite{Cheng_2008} found qualitative agreement with the EQCD
prediction down to about $1.5\,T_{pc}$, while a recent HISQ study with a
continuum extrapolation~\cite{swagatam} found that dimensional reduction sets
in quantitatively only at $T \gtrsim 5\,T_{pc} \approx 780\,\mathrm{MeV}$.
Our work provides complementary HISQ results at two lattice spacings over
the full range $166 \mathrm{MeV}$--$1000\,\mathrm{MeV}$.

\section{Dimensional Reduction: EQCD and MQCD}

The strategy is to exploit the scale hierarchy - (works for $g\ll1$) -  $\pi T \gg gT \gg g^2 T$
by integrating out degrees of freedom at each scale in sequence,
producing a chain of two effective theories (EQCD, MQCD) valid at progressively longer distances.
But for our temperature regions, the value of 'g' is still greater than one, and still we would like to probe the dominance of EQCD and MQCD at these scales.

\paragraph{EQCD.}
Integrating out all non-static ($n \neq 0$) bosonic and fermionic Matsubara
modes, which have masses $\sim 2\pi T$, gives a three-dimensional effective
theory valid for $r \gg 1/(\pi T)$~\cite{Nadkarni1983,Nadkarni1988}:
\begin{equation}
S_3^{E} = \int d^3x \left[
  \tfrac{1}{4}F_{ij}F_{ij}
  + \mathrm{Tr}[D_i,A_0]^2
  + m_E^2\,\mathrm{Tr}[A_0^2]
  + \lambda_A \bigl(\mathrm{Tr}[A_0^2]\bigr)^2
\right],
\label{eq:EQCD}
\end{equation}
where $\partial_\tau A_\mu = 0$, so $A_0^a$ is no longer a gauge field
component but a static adjoint scalar with thermally generated Debye
mass $m_E \sim gT$.
The matching coefficients at two-loop accuracy~\cite{Laine2005} are:
\begin{align}
g_E^2 &= T\!\left[g^2 + \frac{g^4}{4\pi^2}\alpha_3 + \mathcal{O}(g^6)\right],
\quad
m_E^2 = T^2\!\left[g^2\alpha_1 + \frac{g^4}{4\pi^2}\alpha_2\right],
\end{align}
where $g = g(\bar\mu)$ is the $\overline{\mathrm{MS}}$ coupling at
$\bar\mu = (0.5-2)9.1T$, where $\bar\mu$ is scale of minimum sensitivity \cite{Cheng_2008} and the $\alpha_i$ depend on $N_c$ and $n_f$.

\paragraph{MQCD.}
At even higher temperatures $A_0$ is also heavy and can be integrated
out, leaving 3D pure $\mathrm{SU}(3)$ Yang-Mills
theory valid for $r > 1/(g^2T)$:
\begin{equation}
S_3^{M} = \int d^3x\;\tfrac{1}{4}F_{ij}F_{ij},
\end{equation}
with the magnetic coupling matched at two-loop order~\cite{Laine2005}:
\begin{equation}
g_M^2 = g_E^2\!\left[
1 - \frac{1}{48}\frac{g_E^2 N_c}{\pi m_E}
- \frac{17}{4608}\!\left(\frac{g_E^2 N_c}{\pi m_E}\right)^{\!2}
\right].
\label{eq:gM}
\end{equation}
Since $g_M^2$ has dimensions of mass and is the only dimensionful scale
in MQCD, it determines the spatial string tension:
\begin{equation}
\sqrt{\sigma_s} = c\,g_M^2(T),
\label{eq:pred}
\end{equation}
where $c = 0.5530(10)$ is determined from 3D pure-gauge lattice
simulations~\cite{PhysRevD.66.097502}. Further studies found the value of c to be 0.54(1) \cite{Cheng_2008} and 0.566(15) \cite{swagatam}.
This is the central prediction we test.

\section{Lattice Setup}
We use gauge ensembles with the HISQ fermion action at physical 2+1 quark
masses with ($m_l/m_s=1/20)$ and the tree-level Symanzik improved gauge action, generated with the package\\
\texttt{SIMULATeQCD}~\cite{p5} using the RHMC algorithm.
The temperature is $T = 1/(N_\tau a)$, with the lattice spacing set via
$\beta = 6/g_0^2$ and the Sommer parameter from the FLAG Review~\cite{p3}.
We have $N_\tau = 8$ and $N_\tau = 10$ covering $T \in [166,\,1000]$\,MeV with spatial volumes $N_s/N_\tau \geq 4$.

Spatial Wilson loops $W(R,Z)$ are constructed in fixed time-slices for
rectangular loops of transverse extent $R$ and longitudinal extent $Z$, here $Z$ plays role of fictitious time.
For off-axis paths we use the Bresenham algorithm~\cite{PhysRevD.63.074504}, which
provides a lattice approximation to 
-line paths.
The spatial pseudo-potential is:
\begin{equation}
a\,V_s(R) = -\lim_{Z\to\infty}\ln\frac{W(R,Z+1)}{W(R,Z)}.
\end{equation}

\section{Modified HYP Smearing}

The signal-to-noise ratio of spatial Wilson loops degrades with $R$ and $Z$
due to short-distance gauge-field fluctuations.
We suppress these using HYP (Hypercubic Blocking) smearing~\cite{p6},
which replaces each gauge link with a locally averaged fat link while
preserving the lattice symmetries.

In this work we along with 3-level HYP smearing also make use the 2-level HYP scheme, which modifies the original
3-level scheme of Hasenfratz and Knechtli\cite{p6} by ignoring the smearing of the
fictitious time-like links ($U_{i,\rho}$--- bare link at location \textit{i} in direction $\rho$).
The construction proceeds in two steps.
At the final level, the blocked link (black) $V_{i,\mu;\rho}$ is constructed via SU(3) projected modified APE smearing :
\begin{equation}
V_{i,\nu;\rho} = \mathrm{Proj}_{SU(3)}\!\left[
  (1-\alpha_2)\,U_{i,\nu\ne\rho}
  + \frac{\alpha_2}{4}\sum_{\pm\mu\neq\nu,\rho}
    \tilde{V}_{i,\mu;\nu,\rho}\,\tilde{V}_{i+\mu,\nu;\mu,\rho}\,
    \tilde{V}^\dagger_{i+\nu,\mu;\nu,\rho}
\right],
\label{eq:HYP2}
\end{equation}

\begin{figure}[H]
  \centering
  \includegraphics[width=0.60\linewidth]{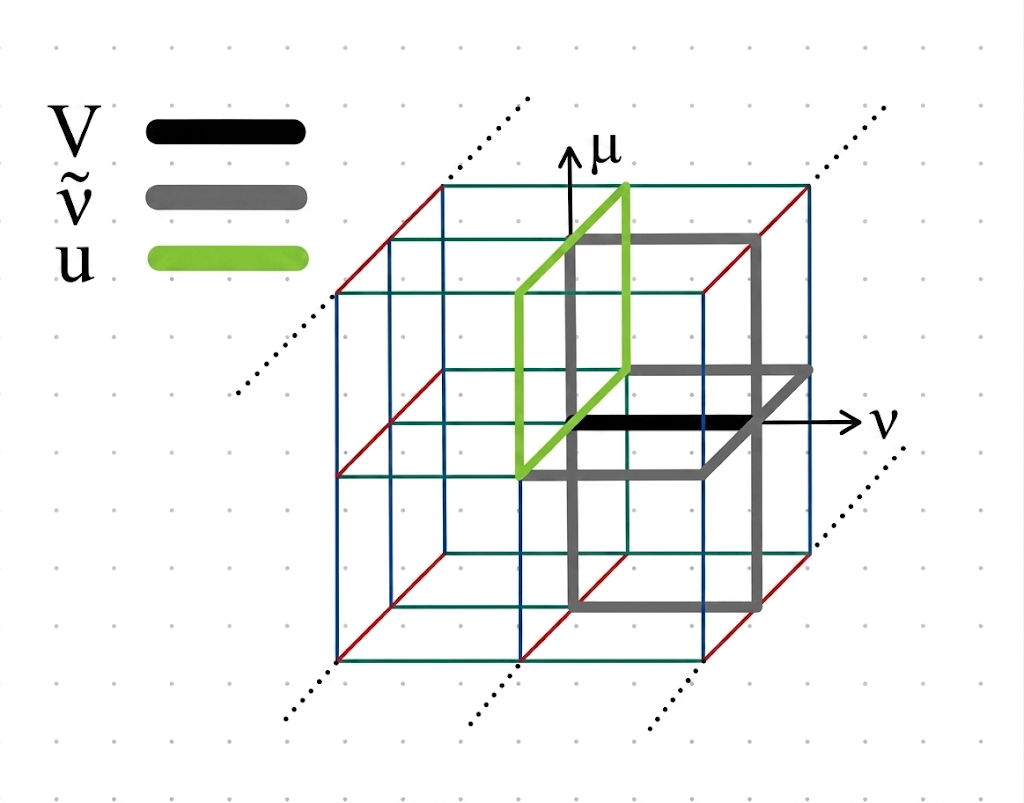}
  \caption{3D spatial lattice showing the 2-level HYP smearing construction.
  The black link $V_{i,\mu}$ is the final smeared link which is made from modified APE smearing of fat links $\tilde{V}$, where further these fat links $\tilde{V}_{i,\mu;\nu,\rho}$ are made from modified APE blocking of green bare links ${U}_{i,\eta\ne(\nu,\rho)}$.
  }
  \label{fig:hyp}
\end{figure}

Here, the $\rho$ is fictitious time direction (any one of spatial direction) and $\nu$ is direction of original link to be smeared. 
The $\nu,\rho$ in $\tilde{V}_{i,\mu;\nu,\rho}$ implies that the fat link at site \textit{i} in the direction $\mu$ is not decorated with staples extending in direction $\nu,\rho$. This construction ensures that we only include the links within the hypercube attached to original link $U_{i,\nu}$.
Where, the decorated fat link $\tilde{V}_{i,\mu;\nu,\rho}$ (grey) is built from the bare links $U_{i,\eta}$(green):
\begin{equation}
\tilde{V}_{i,\mu;\nu,\rho} = \mathrm{Proj}_{SU(3)}\!\left[
  (1-\alpha_3)\,U_{i,\mu\ne\rho}
  + \frac{\alpha_3}{2}\sum_{\pm\eta\neq\nu,\mu,\rho}
    U_{i,\eta}\,U_{i+\hat\eta,\mu}\,U^\dagger_{i+\hat\mu,\eta}
\right].
\label{eq:HYP1}
\end{equation}

with $\alpha_2 = 0.6$ and $\alpha_3 = 0.3$. $U_{i,\mu\ne\rho}$ implies the bare link at location \textit{i} in direction $\mu\ne\rho$. Here in our construction, euclidean time links are treated like other spatial links.
Figure~\ref{fig:hyp} shows this construction schematically.
We apply 3 and 5 smearing steps with both 2-level and 3-level HYP,
giving four smearing schemes in total, and use their mutual agreement
as a diagnostic for systematic uncertainty.

\section{Extracting the Spatial String Tension}

\subsection*{Effective mass fits and the pseudo-potential}

\begin{figure}[H]
  \centering
  \begin{subfigure}[b]{0.48\linewidth}
    \includegraphics[width=\linewidth]{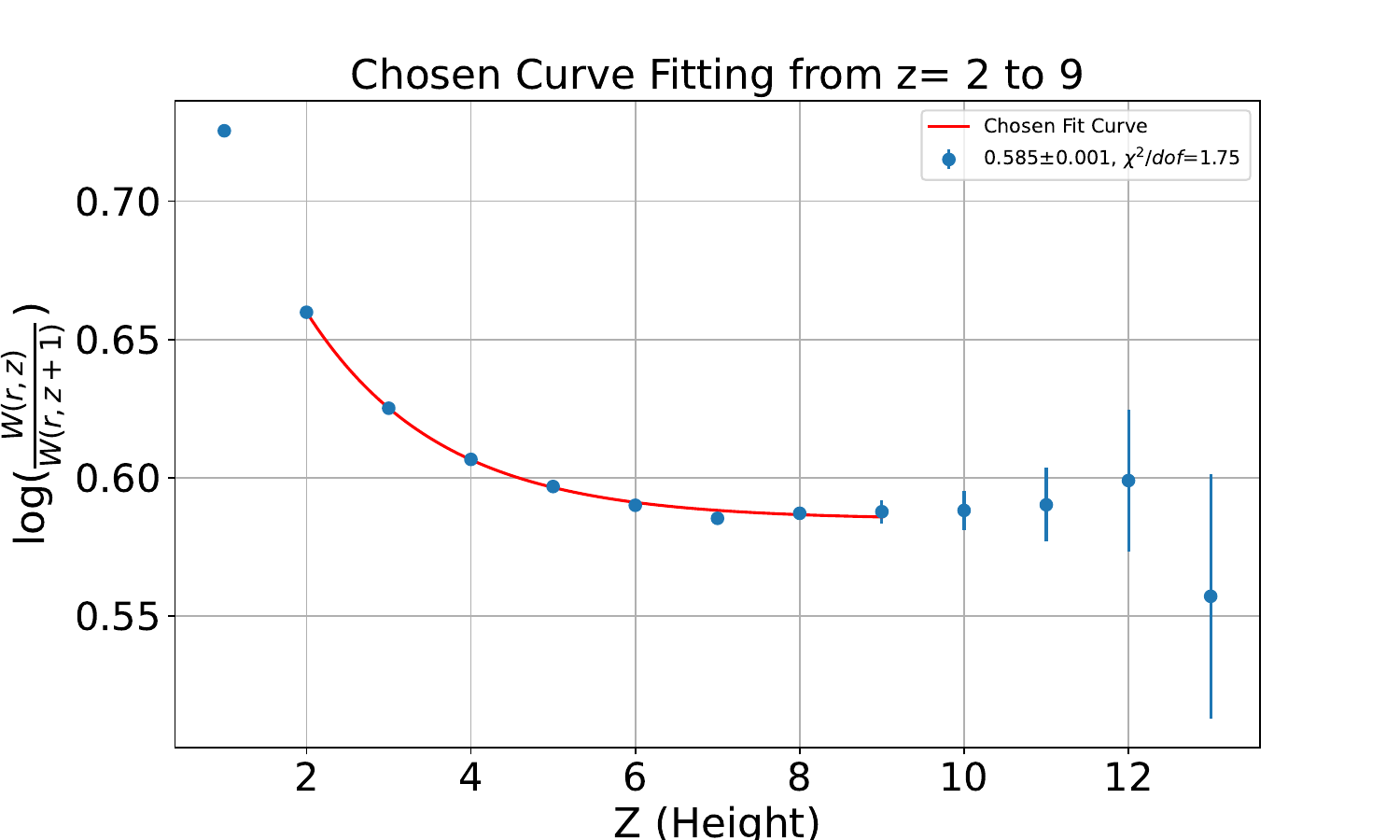}
    \caption{2-level HYP, 3 steps.}
  \end{subfigure}\hfill
  \begin{subfigure}[b]{0.48\linewidth}
    \includegraphics[width=\linewidth]{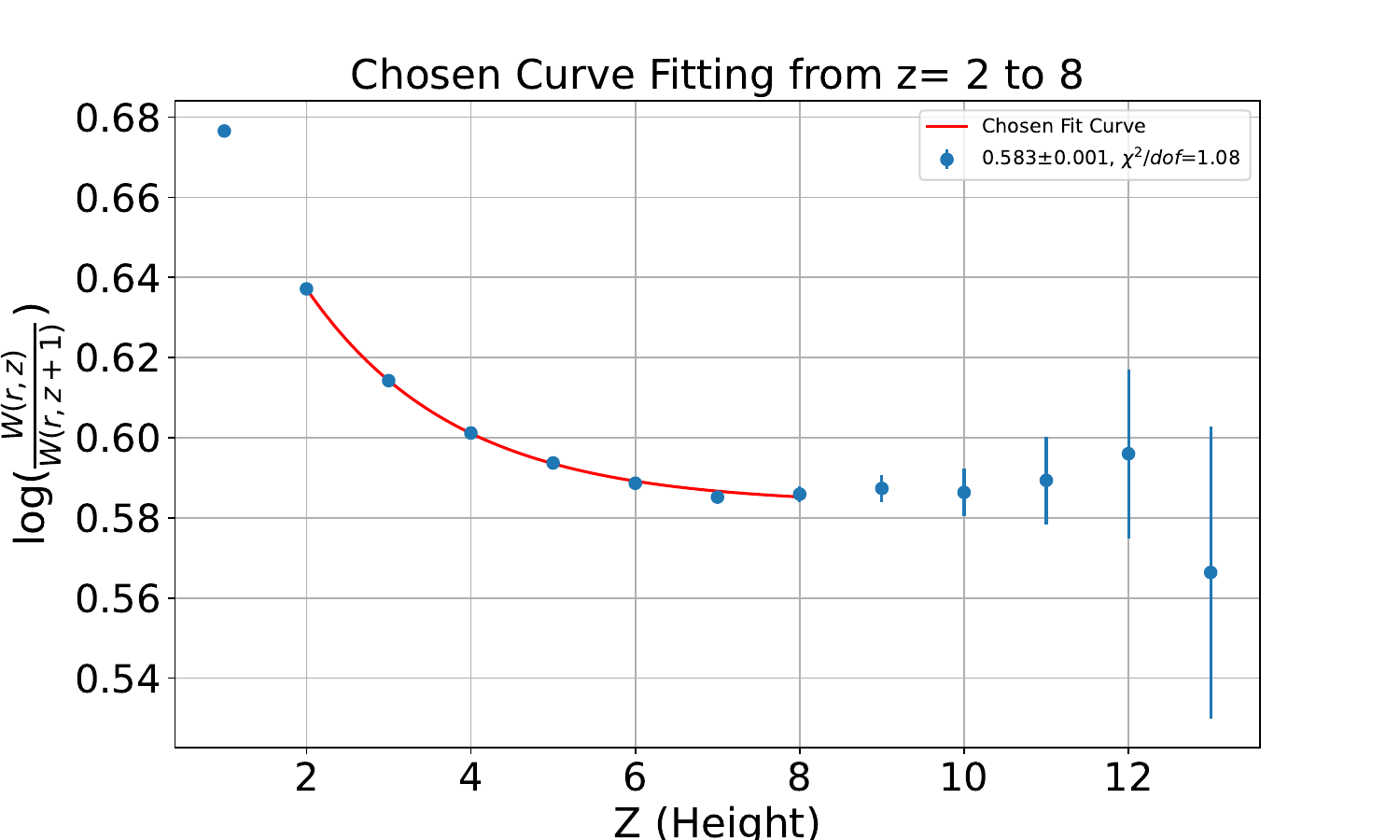}
    \caption{2-level HYP, 5 steps.}
  \end{subfigure}
  \vspace{0.5em}
  \begin{subfigure}[b]{0.48\linewidth}
    \includegraphics[width=\linewidth]{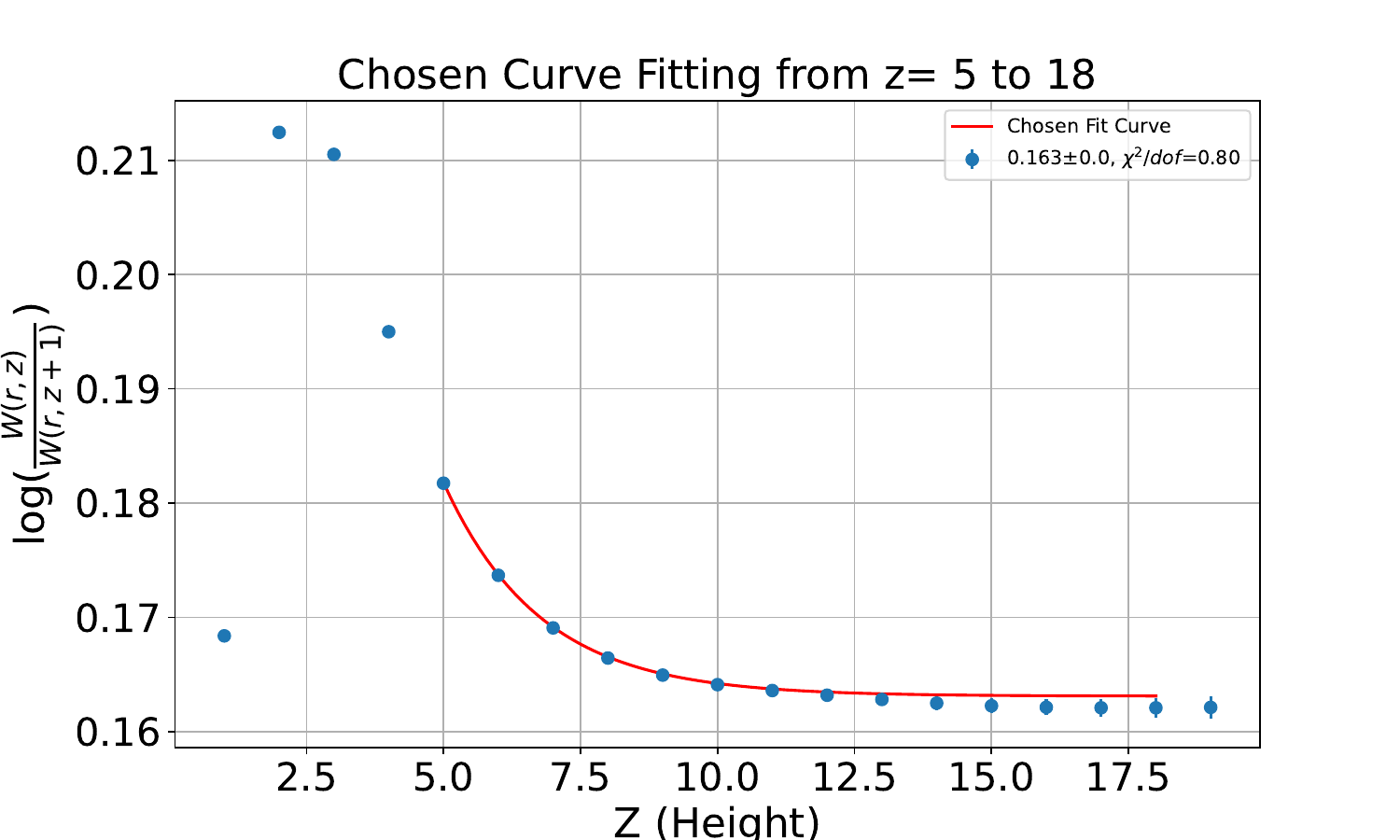}
    \caption{3-level HYP, 3 steps.}
  \end{subfigure}\hfill
  \begin{subfigure}[b]{0.48\linewidth}
    \includegraphics[width=\linewidth]{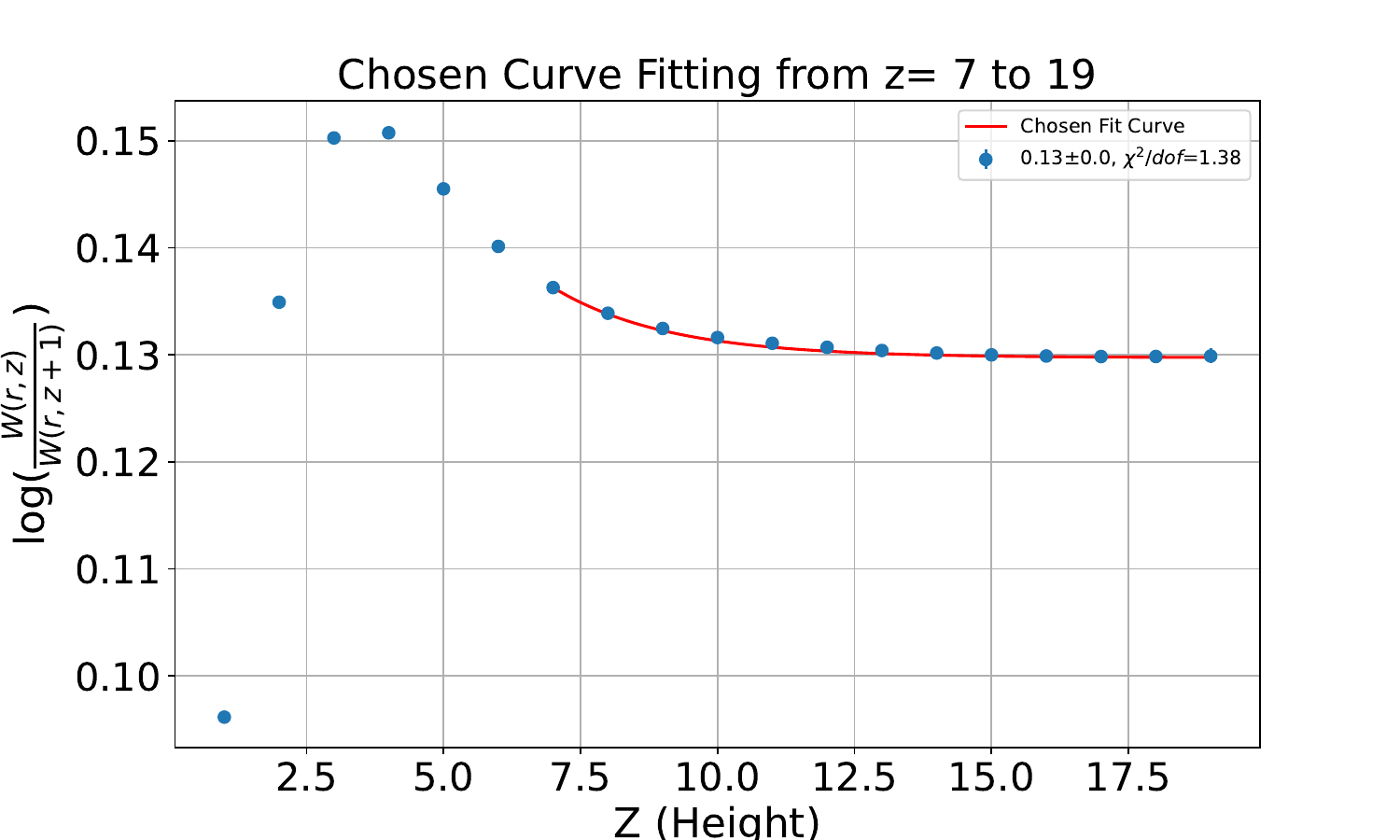}
    \caption{3-level HYP, 5 steps.}
  \end{subfigure}
  \caption{Effective mass $\log[W(r,z)/W(r,z+1)]$ vs $Z$ for $\beta = 8.570, N_\tau=10$, at $r/a=7.2111$,
  $T \approx 924\,\mathrm{MeV}$, for all four smearing schemes.
  The red curve is the fit $V_s(R) + b\,e^{-V_1 Z}$.
  The ground-state value $V_s(R)$ and $\chi^2/\mathrm{dof}$ are in the legend.}
  \label{fig:effmass}
\end{figure}

The pseudo-potential $V_s(R)$ cannot be read directly from $W(R,Z)$ at
finite $Z$ because excited states contaminate the signal at small $Z$.
We therefore fit the effective mass to the one-excited-state ansatz:
\begin{equation}
m(Z,R) = \ln\frac{W(R,Z)}{W(R,Z+1)} = V_s(R) + b\,e^{-V_1 Z},
\label{eq:effmass}
\end{equation}
and extract the ground-state value $V_s(R)$ from the plateau.
Figure~\ref{fig:effmass} shows representative effective mass fits for
$\beta = 8.570$ ($T \approx 924\,\mathrm{MeV}$, $N_\tau = 10$) for all
four smearing schemes.
More smearing steps bring the plateau in at smaller $Z$, as expected,
since they suppress the short-distance excited-state contributions.
Gauge configurations are separated by 10 RHMC steps; autocorrelation
times were below one for all ensembles. Since these fits were obtained from minimising correlated $\chi^2$, the
statistical errors quoted here are fit errors; a full jackknife analysis
over configurations is part of the ongoing work.

\subsection*{Renormalisation and Cornell fit}

The pseudo-potential carries an additive ultraviolet divergence from the
self-energy of the static colour sources.
We remove it by fixing the renormalisation constant $C_Q(\beta)$ through
the zero-temperature quark--antiquark potential~\cite{p8}:
\begin{equation}
C_Q \;\text{set by}\;
\begin{cases}
V(r_0) = 0.954/r_0 & \beta \leq 6.488,\\
V(r_1) = 0.2065/r_1 & \beta > 6.488,
\end{cases}
\end{equation}
giving $V^{\mathrm{ren}} = V^{\mathrm{bare}} + 2C_Q$.
We then fit $V^{\mathrm{ren}}(r)$ to the Cornell form:
\begin{equation}
V_s(r) = \sigma_s\,r - \frac{\alpha}{r} + c,
\label{eq:cornell}
\end{equation}
with $\alpha$ fixed to the Luscher-term value:
\begin{equation}
\alpha = \begin{cases}
\pi/12 & T < 5\,T_{pc},\\
\pi/24 & T \geq 5\,T_{pc}.
\end{cases}
\end{equation}
This gives a two-parameter fit in $\sigma_s$ and $c$.
We set $r_{min}>1/(gT)$ and three-parameter fits with
free $\alpha$ are used as a systematic check.

\section{Results}
\label{sec:results}

\subsection*{Spatial potential}

Figure~\ref{fig:potential} shows the renormalised spatial potential $V(r)$
for all ensembles using 2-level HYP smearing with 3 smearing steps at
$N_\tau = 8$ and $N_\tau = 10$.
There is a clear linearly rising signal across all temperatures, and the
Cornell fits have $\chi^2_{\mathrm{red}}$ close to unity in most cases.
The slope grows with temperature, consistent with $\sigma_s \propto g^2(T)\,T$
as expected from Eq.~\eqref{eq:pred}.
\begin{figure}[H]
  \centering
  \begin{subfigure}[b]{0.5\linewidth}
    \includegraphics[width=\linewidth]{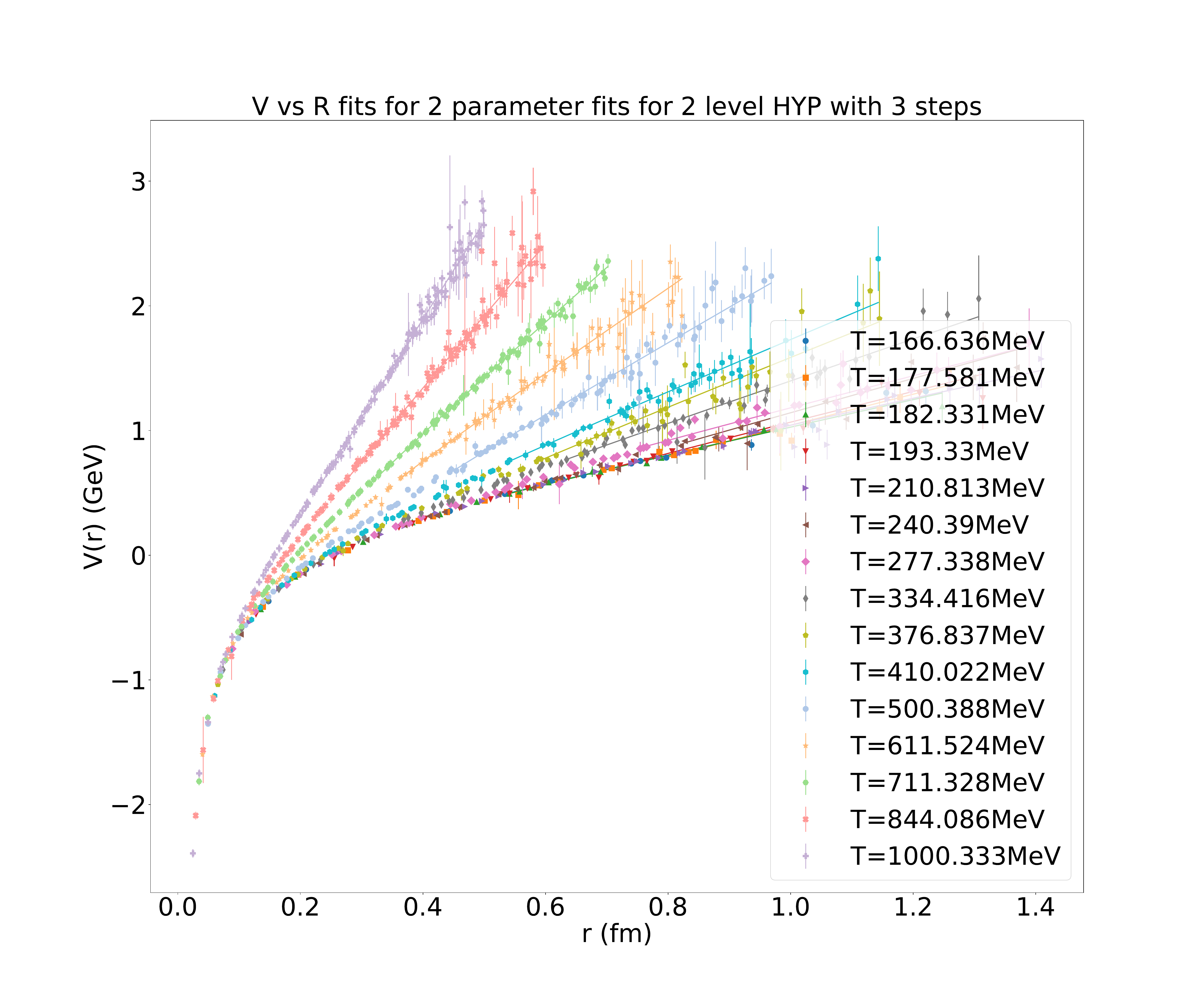}
    \caption{2-level HYP, 3 steps. ($N_\tau=8$)}
  \end{subfigure}\hfill
  \begin{subfigure}[b]{0.5\linewidth}
    \includegraphics[width=\linewidth]{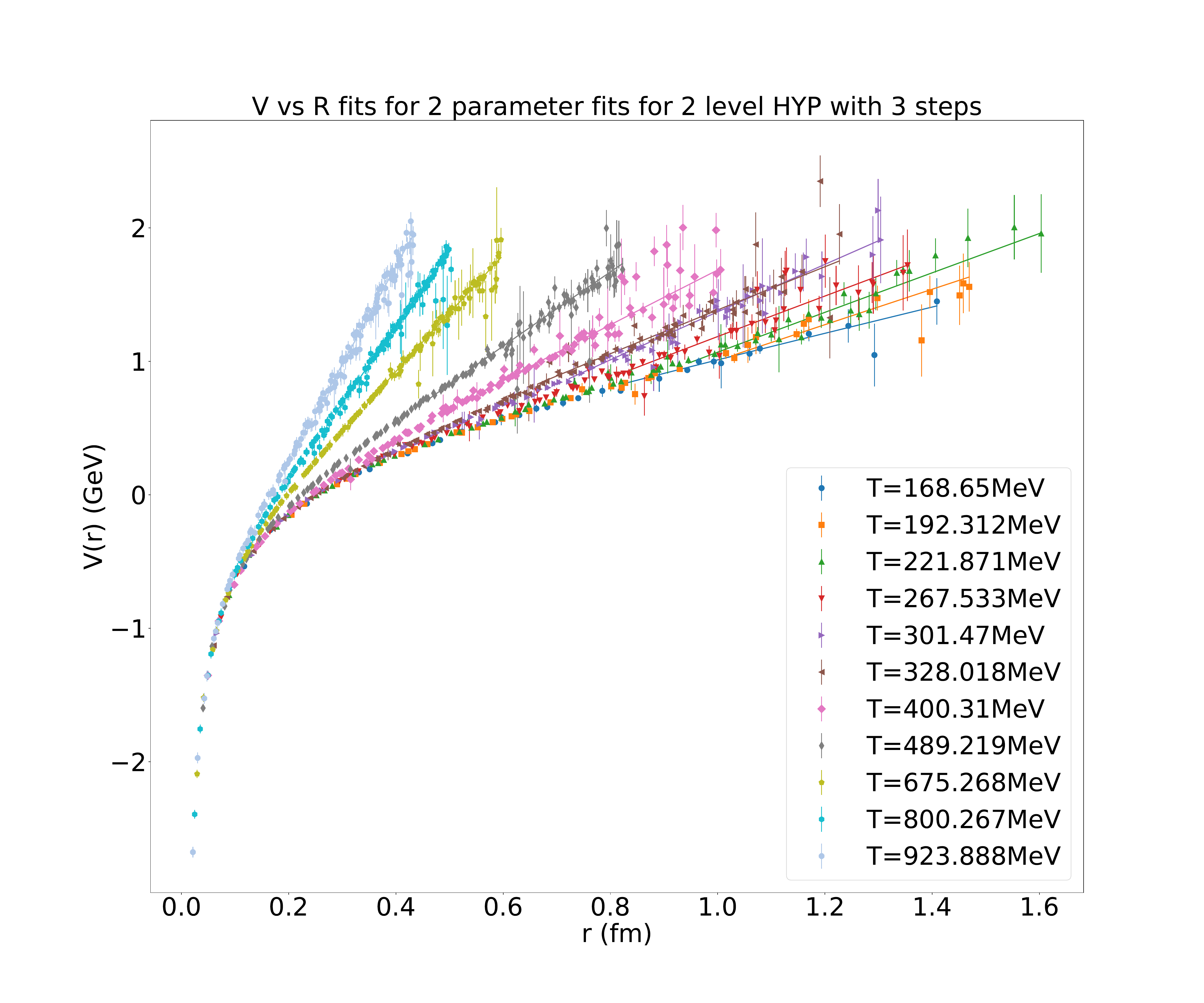}
    \caption{2-level HYP, 3 steps.($N_\tau=10$)}
  \end{subfigure}
\caption{Renormalized spatial potential $V(r)$ vs $r\,[\mathrm{fm}]$ for all
  ensembles using 2-level HYP smearing with 3 smearing steps,
  at $N_\tau = 8$ (left) and $N_\tau = 10$ (right).
  Solid lines are Cornell fits (Eq.~\eqref{eq:cornell}); the legend gives
 temperature in MeV.}

\end{figure}

%
  \label{fig:potential}

\subsection*{Comparison with MQCD}

\begin{figure}[H]
  \centering
  \includegraphics[width=\linewidth]{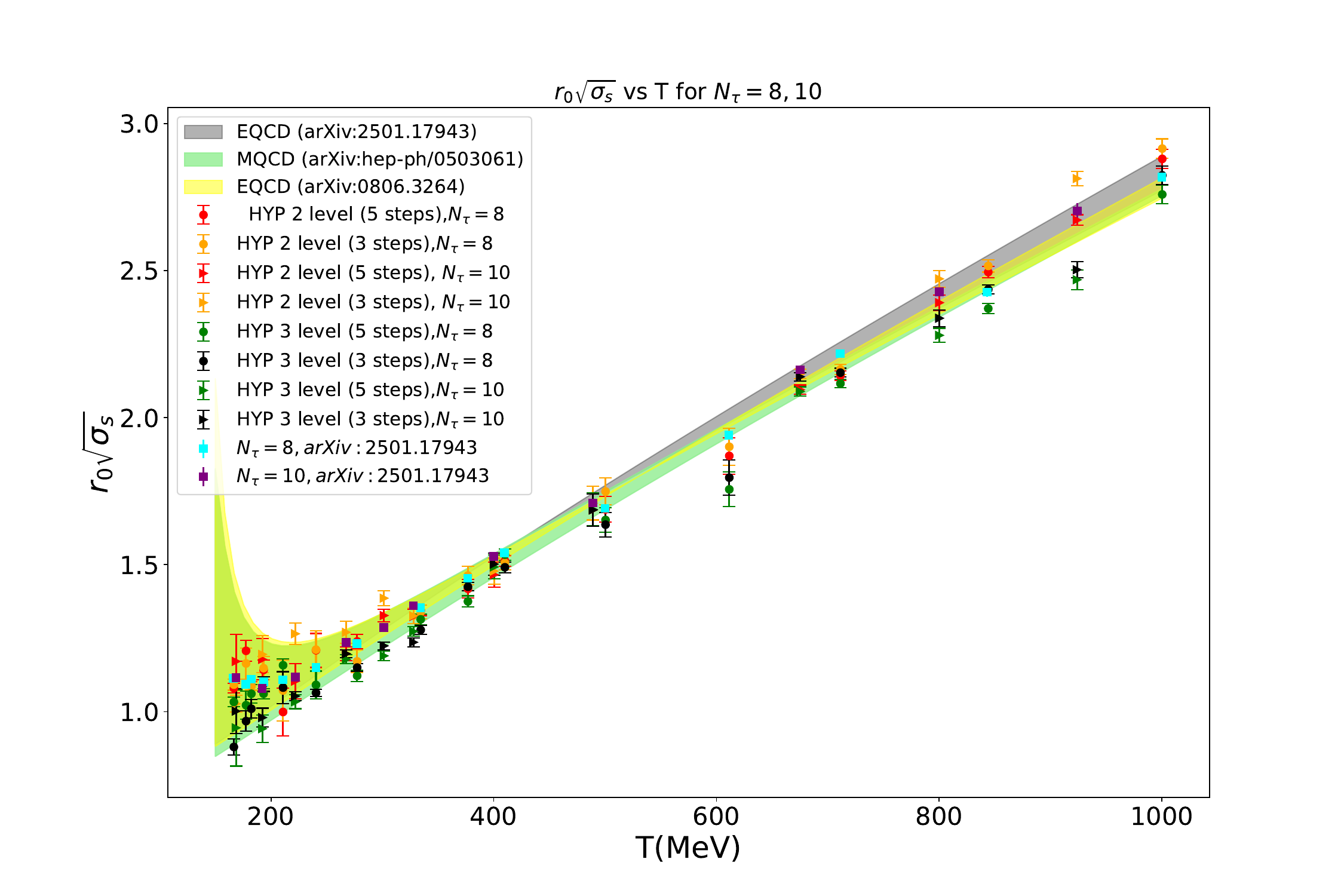}
  \caption{$r_0\sqrt{\sigma_s}$ vs temperature for $N_\tau = 8$ and $N_\tau = 10$,
  with $g$ evaluated at 5-loop \cite{Herren_2018} order at $\bar\mu = (0.5-2)9.1T$. Where $\bar \mu$ is minimum sensitive energy scale \cite{Cheng_2008}.
  Results for all four smearing schemes: 2-level and 3-level HYP,
  each with 3 and 5 steps with (red,orange) and (green,black) symbols respectively.
  Grey band: EQCD prediction from Ref.~\cite{swagatam};
  green band: MQCD prediction from Ref.~\cite{Laine2005};
  yellow band: EQCD prediction from Ref.~\cite{Cheng_2008}.
  Cyan/purple: $N_\tau = 8, 10$ results from Ref.~\cite{swagatam}.
  The four schemes do not agree within error bars, indicating an
  uncontrolled systematic that needs to be resolved.}
  \label{fig:main}
\end{figure}

Figure~\ref{fig:main} shows the main result: $r_0\sqrt{\sigma_s}$ as a
function of temperature for all four smearing schemes at both $N_\tau$ values,
compared with the EQCD and MQCD predictions using 5-loop running coupling
$g(\bar\mu)$ at $\bar\mu = (0.5 - 2)*9.1* T$.
The data rise monotonically with temperature, as expected from
$\sqrt{\sigma_s} \propto g_M^2(T)$.
At high temperatures ($T \gtrsim 600\,\mathrm{MeV}$) the points approach
the MQCD band, consistent with dimensional reduction setting in around
$T \sim 5\,T_{pc}$.

However, the results from the different smearing schemes do not agree with
each other within statistical error bars for some of the temperatures, revealing an uncontrolled
systematic in the current analysis.
A naive continuum extrapolation using $N_\tau = 8$ and $10$ was attempted
but could not be completed, since the two sets of data disagree at several
temperatures beyond statistical errors.
The root cause --- whether it lies in the treatment of excited-state contamination or fit stability tests ---
is being investigated, and the analysis is being revisited.

\section{Conclusion}

Since this is work in progress, strong conclusions cannot be drawn at this stage.
We have calculated the spatial string tension in 2+1 flavour QCD over
$166$--$1000\,\mathrm{MeV}$ at $N_\tau = 8$ and $10$, and compared
with the MQCD prediction.

The extracted $r_0\sqrt{\sigma_s}$ rises monotonically with temperature
and approaches the MQCD band at high $T$.
The values of $\sigma_s$ were obtained using the maximum fit range
rather than a plateau, though the plateau was not observed to be far from
the maximum-range result.
The results from the four smearing schemes do not agree within error bars,
and a continuum extrapolation was not achievable.
A precise determination of $\alpha$ and a full continuum extrapolation
remain to be done.
The calculations are being revisited and will be
published in the coming months.

\section*{Acknowledgments}

We thank Sayantan Sharma and Swagatam Tah for useful discussions
during the conference.
Computations for Spatial Wilson Loop and Polyakov Loop (for calculation of $C_Q$ for different smear steps) were performed in part at the Center for High Energy Physics, IISc Bangalore, using SIMULATeQCD package \cite{p5}
. PP was supported by U.S. Department
of Energy, Office of Science, Office of Nuclear Physics under Contract
No.DE-SC0012704.

\bibliographystyle{unsrturl}
\bibliography{reference}

\begin{thebibliography}{10}

\bibitem{HotQCD2019}
A.~Bazavov et~al.
\newblock Chiral crossover in {QCD} at zero and non-zero chemical potentials.
\newblock {\em Phys.\ Lett.\ B}, 795:15--21, 2019.
\newblock \href {https://arxiv.org/abs/1812.08235} {\path{arXiv:1812.08235}}.

\bibitem{Linde1980}
A.~D. Linde.
\newblock Infrared problem in the thermodynamics of the {Yang-Mills} gas.
\newblock {\em Phys.\ Lett.\ B}, 96:289--292, 1980.
\newblock \href {https://doi.org/10.1016/0370-2693(80)90769-8}
  {\path{doi:10.1016/0370-2693(80)90769-8}}.

\bibitem{Cheng_2008}
M.~Cheng, S.~Datta, J.~van~der Heide, K.~Huebner, F.~Karsch, O.~Kaczmarek,
  E.~Laermann, J.~Liddle, R.~D. Mawhinney, C.~Miao, P.~Petreczky, K.~Petrov,
  C.~Schmidt, W.~Soeldner, and T.~Umeda.
\newblock The spatial string tension and dimensional reduction in qcd.
\newblock {\em Physical Review D}, 78(3), August 2008.
\newblock URL: \url{http://dx.doi.org/10.1103/PhysRevD.78.034506}, \href
  {https://doi.org/10.1103/physrevd.78.034506}
  {\path{doi:10.1103/physrevd.78.034506}}.

\bibitem{swagatam}
Dibyendu Bala, Olaf Kaczmarek, Peter Petreczky, Sayantan Sharma, and Swagatam
  Tah.
\newblock The spatial string tension and its effects on screening correlators
  in a thermal qcd plasma, 2025.
\newblock URL: \url{https://arxiv.org/abs/2501.17943}, \href
  {https://arxiv.org/abs/2501.17943} {\path{arXiv:2501.17943}}.

\bibitem{Nadkarni1983}
S.~Nadkarni.
\newblock Dimensional reduction in hot {QCD}.
\newblock {\em Phys.\ Rev.\ D}, 27:917--931, 1983.
\newblock \href {https://doi.org/10.1103/PhysRevD.27.917}
  {\path{doi:10.1103/PhysRevD.27.917}}.

\bibitem{Nadkarni1988}
S.~Nadkarni.
\newblock Non-abelian debye screening. {II}. {T}he singlet potential.
\newblock {\em Phys.\ Rev.\ D}, 38:3287--3294, 1988.
\newblock \href {https://doi.org/10.1103/PhysRevD.38.3287}
  {\path{doi:10.1103/PhysRevD.38.3287}}.

\bibitem{Laine2005}
M.~Laine and Y.~Schr\"{o}der.
\newblock Two-loop {QCD} gauge coupling at high temperatures.
\newblock {\em JHEP}, 03:067, 2005.
\newblock \href {https://arxiv.org/abs/hep-ph/0503048}
  {\path{arXiv:hep-ph/0503048}}.

\bibitem{PhysRevD.66.097502}
Biagio Lucini and Michael Teper.
\newblock $\mathrm{SU}(n)$ gauge theories in $2+1$ dimensions: Further results.
\newblock {\em Phys. Rev. D}, 66:097502, Nov 2002.
\newblock URL: \url{https://link.aps.org/doi/10.1103/PhysRevD.66.097502}, \href
  {https://doi.org/10.1103/PhysRevD.66.097502}
  {\path{doi:10.1103/PhysRevD.66.097502}}.

\bibitem{p5}
Lukas Mazur, Dennis Bollweg, David~A. Clarke, Luis Altenkort, Olaf Kaczmarek,
  Rasmus Larsen, Hai-Tao Shu, Jishnu Goswami, Philipp Scior, Hauke Sandmeyer,
  Marius Neumann, Henrik Dick, Sajid Ali, Jangho Kim, Christian Schmidt, Peter
  Petreczky, and Swagato Mukherjee.
\newblock Simulateqcd: A simple multi-gpu lattice code for qcd calculations.
\newblock {\em Computer Physics Communications}, 300:109164, July 2024.
\newblock URL: \url{http://dx.doi.org/10.1016/j.cpc.2024.109164}, \href
  {https://doi.org/10.1016/j.cpc.2024.109164}
  {\path{doi:10.1016/j.cpc.2024.109164}}.

\bibitem{p3}
Y.~Aoki, T.~Blum, S.~Collins, L.~Del Debbio, M.~Della Morte, P.~Dimopoulos,
  X.~Feng, M.~Golterman, Steven Gottlieb, R.~Gupta, G.~Herdoiza, P.~Hernandez,
  A.~Jüttner, T.~Kaneko, E.~Lunghi, S.~Meinel, C.~Monahan, A.~Nicholson,
  T.~Onogi, P.~Petreczky, A.~Portelli, A.~Ramos, S.~R. Sharpe, J.~N. Simone,
  S.~Sint, R.~Sommer, N.~Tantalo, R.~Van de~Water, A.~Vaquero, U.~Wenger, and
  H.~Wittig.
\newblock Flag review 2024, 2025.
\newblock URL: \url{https://arxiv.org/abs/2411.04268}, \href
  {https://arxiv.org/abs/2411.04268} {\path{arXiv:2411.04268}}.

\bibitem{PhysRevD.63.074504}
Bram Bolder, Thorsten Struckmann, Gunnar~S. Bali, Norbert Eicker, Thomas
  Lippert, Boris Orth, Klaus Schilling, and Peer Ueberholz.
\newblock High precision study of the $q\overline{Q}$ potential from wilson
  loops at large distances.
\newblock {\em Phys. Rev. D}, 63:074504, Mar 2001.
\newblock URL: \url{https://link.aps.org/doi/10.1103/PhysRevD.63.074504}, \href
  {https://doi.org/10.1103/PhysRevD.63.074504}
  {\path{doi:10.1103/PhysRevD.63.074504}}.

\bibitem{p6}
Anna Hasenfratz and Francesco Knechtli.
\newblock {Flavor symmetry and the static potential with hypercubic blocking}.
\newblock {\em Phys. Rev. D}, 64:034504, 2001.
\newblock \href {https://arxiv.org/abs/hep-lat/0103029}
  {\path{arXiv:hep-lat/0103029}}, \href
  {https://doi.org/10.1103/PhysRevD.64.034504}
  {\path{doi:10.1103/PhysRevD.64.034504}}.

\bibitem{p8}
A.~Bazavov, N.~Brambilla, H.~T. Ding, P.~Petreczky, H.~P. Schadler, A.~Vairo,
  and J.~H. Weber.
\newblock Polyakov loop in 2+1 flavor qcd from low to high temperatures, 2016.
\newblock URL: \url{https://arxiv.org/abs/1603.06637}, \href
  {https://arxiv.org/abs/1603.06637} {\path{arXiv:1603.06637}}, \href
  {https://doi.org/10.1103/PhysRevD.93.114502}
  {\path{doi:10.1103/PhysRevD.93.114502}}.

\bibitem{Herren_2018}
Florian Herren and Matthias Steinhauser.
\newblock Version 3 of rundec and crundec.
\newblock {\em Computer Physics Communications}, 224:333–345, March 2018.
\newblock URL: \url{http://dx.doi.org/10.1016/j.cpc.2017.11.014}, \href
  {https://doi.org/10.1016/j.cpc.2017.11.014}
  {\path{doi:10.1016/j.cpc.2017.11.014}}.

\end{thebibliography}

\end{document}